\documentclass{ragtime} 

\title[Collapsing massive stars and their EM transients]%
      {Collapsing massive stars with self-gravity \\ and their electromagnetic transients}

\author[A. Janiuk, 
        N. Shahamat   
        and D. Kr\'ol]
       {Agnieszka Janiuk\at[]{1,a} 
        Narjes Shahamat\at{2} 
        and Dominika Kr\'ol \at{3}\\
        \ins{1}Center for Theoretical Physics,
        Polish Academy of Sciences\splitins[1]
        Al. Lotnik\'ow 32/46, 02-668, Warsaw,
        Poland \\
        \ins{2} Department of Physics, School of Science\splitins[1]%
        Ferdowsi University, Mashhad, Iran\\
        \ins{3} Astronomical Observatory, Jagiellonian University\splitins[1]%
        Krak\'ow, Poland\\
        \ins{a}\Email{agnes@cft.edu.pl}} 

\coentry{A. Janiuk, N. Shahamat, and D. Kr\'ol}%
       {Collapsing stars\par
        from long GRBs}

\begin{document}

\begin{abstract}
We investigate the fate of a collapsing stellar core, which is the final state of evolution of a massive, rotating star
of a Wolf-Rayet type. Such stars explode as type I b/c supernovae, which have been observed in association with long gamma ray bursts
(GRBs).
The core of the star is potentially forming a black hole, which is embedded in a dense, rotating, and possibly highly magnetized envelope.
We study the process of collapse using General Relativistic MHD simulations, and we account for the growth of the black hole
 mass and its spin, as well as related evolution of the spacetime metric. 
We find that some particular configurations of the initial black hole spin, the content of angular momentum in the stellar core, and the magnetic field configuration and its strength, are favored for producing a bright electromagnetic transient (i.e., a gamma ray burst). On the other hand, most of the typical configurations studied in our models do not lead to a transient electromagnetic explosion and will end up in a direct collapse, 
accompanied by some residual variability induced by changing accretion rate.
We also study the role of self-gravity in the stellar core and quantify the relative strength of the interfacial instabilities, such as Self-Gravity Interfacial (SGI) instability and Rayleigh-Taylor (RT), which may account for the production of an inhomogeneous structure, including spikes and bubbles, through the inner radii of the collapsing core (inside $\sim 200~r_{g}$).
We find that in self-gravitating collapsars the RT modes cannot grow efficiently. We also conclude that transonic shocks are formed in the collapsing envelope, but they are weaker in magnetized stars.  
\end{abstract}

\begin{keywords}
Accretion--black hole physics~-- gravitation -- magnetohydrodynamics --
massive stars -- gamma ray bursts
\end{keywords}

\section{Introduction}\label{intro}


Long gamma ray bursts (GRBs) originate from the collapse of massive, rotating stars. Some of the GRBs exhibit much stronger variability patterns in the prompt GRB emission  than the usual stochastic variations. We discuss the mechanisms
of these variations in the frame of self-gravitating collapsar model.

Our computations confirm that gravitational instability can account for flaring activity in GRBs and the variations in their prompt emission. Rapid variability detected in the brightest GRBs, most likely powered by spinning black holes, is consistent with the self-gravitating collapsar model, where the
density inhomogeneities
are formed. The transonic fshocks may also appear, but their
effect should be weakened by magnetic field.

We calculate the time evolution of the collapsing massive star using the General Relativisitic Magneto-Hydrodynamic (GR MHD) scheme.
We have developed a new version of the code HARM-METRIC, upgraded from that presented in \citet{Janiuk2018}.
The evolution of the space-time Kerr metric is accounted for by the increasing mass and changing spin of the black hole.
We added also the new terms, that describe the self-gravity of the star
and are changing at every time-step during dynamical simulation.

In our formulation, the black hole has been already formed
in the centre of the collapsing stellar core and its initial mass in of $3 M_{\odot}$.
Our computational grid size is of $1000 ~ r_{g}$, which makes it smaller
than a compact C-O core of a Wolf-Rayet star or a presupernova.
Thereofre, our model is compact enough to address the problem of self-gravitating gas close to the horizon of a newly formed black hole,
but we do not address any prior or ongoing supernova explosion.

Depending on the rotation of the star, the ultimate outcome might be either a direct collapse or the formation of a mini-disc inside the core, that is, a collapsar
which may lead to an electromagnetic transient.
At the onset of the GRB, the collapsar consists of a black hole, stellar envelope composed of accreting shells with decreasing density, and rotationally supported disc formed at the equatorial region.
At any chosen radius above the horizon, the gas is subject to gravity force induced by the Kerr black hole, the centrifugal force due to envelope rotation, and in addition, it feels the perturbative force due to the self-gravity of the matter, enclosed within a given radius.

\section{Numerical code and setup}\label{code}

We use the general relativistic MHD code called high-accuracy relativistic magnetohydrodynamics (HARM), originally published by \cite{Gammie2003} and further developed by various groups. Our code version, HARM-METRIC, includes the Kerr metric evolution, as first described in \citet{Janiuk2018}.

The code introduces a conservative, shock-capturing scheme with low numerical viscosity to solve the hyperbolic system of partial differential equations of GR MHD.
The numerical scheme uses the plasma energy-momentum tensor, with contributions from matter (gas) and electromagnetic field. For the GR MHD evolution, two fundamental equations are solved for the mass and energy-momentum conservation.
\begin{equation}
(\rho u^{\mu})_{;\mu}=0;~~~~~~~~T_{\nu;\mu}^{\mu}=0.
\end{equation}
\begin{equation}
T_{(m)}^{\mu \nu}=\rho hu^{\mu}u^{\nu}+pg^{\mu \nu}.
\end{equation}
\begin{equation}
T_{(em)}^{\mu \nu}=b^{k}b_{k} hu^{\mu}u^{\nu}+\frac{1}{2} b^{k}b_{k}g^{\mu \nu}-b^{\mu}b^{\nu}.
\end{equation}
\begin{equation}
T^{\mu \nu}=T_{(m)}^{\mu \nu}+T_{(em)}^{\mu \nu}.
\end{equation}
An additional constraint is given by the Equation of State (EOS). In the current project, we used analytic form of adiabatic EOS that relates gas pressure with density. This scales with the power of 4/3, as adequate for a relativistic gas of degenerate particles.

\begin{equation}
p= K \rho^{\gamma}; ~~~ \gamma={4 \over 3}
\end{equation}

The HARM code works in dimensionless units of G = c = 1. Conversion coefficients can be found in \ref{Tab:1}, where the black hole of 3 Solar masses is assumed. Notice that in the plots below, we use geometric unit to express distance, while physical units are used to express time.
  \begin{table}
\begin{center}
  \begin{tabular}{ccc}
\hline
Physical  & Geometrical  & cgs  \\
~ quantity& ~ units      & ~ units \\
\hline
Length & \boldmath$r_g=\frac{GM}{c^2}$ & $4.44 \times 10^{5}$ ~ cm \\ 
Time & \boldmath$T_{unit} = \frac{r_g}{c}$ & $1.38 \times 10^{-5}$ ~ s \\ 
\hline
  \end{tabular}
  \caption{\label{Tab:1}
    Conversion units between numerical code and physical scale of the collapsar.
  }
\end{center}
  \end{table}

\subsection{Initial conditions}

Initial conditions for our collapsing stellar core are given by quasi-spherical distribution of gas endowed with small angular momentum, concentrated at the equatorial plane \citep{KrolJaniuk2021}. The distributions of density and radial velocity are obtained from the Bondi solution, integrated numerically below and above the sonic point. The sonic point is a parameter of our model, and here it is assumed at $80 r_{g}$. Below this point, matter flows into black hole supersonically, and reaches the speed of light at the horizon.

\begin{figure}[p]
\begin{center}
  \includegraphics[width=0.45\linewidth,height=0.35\linewidth]{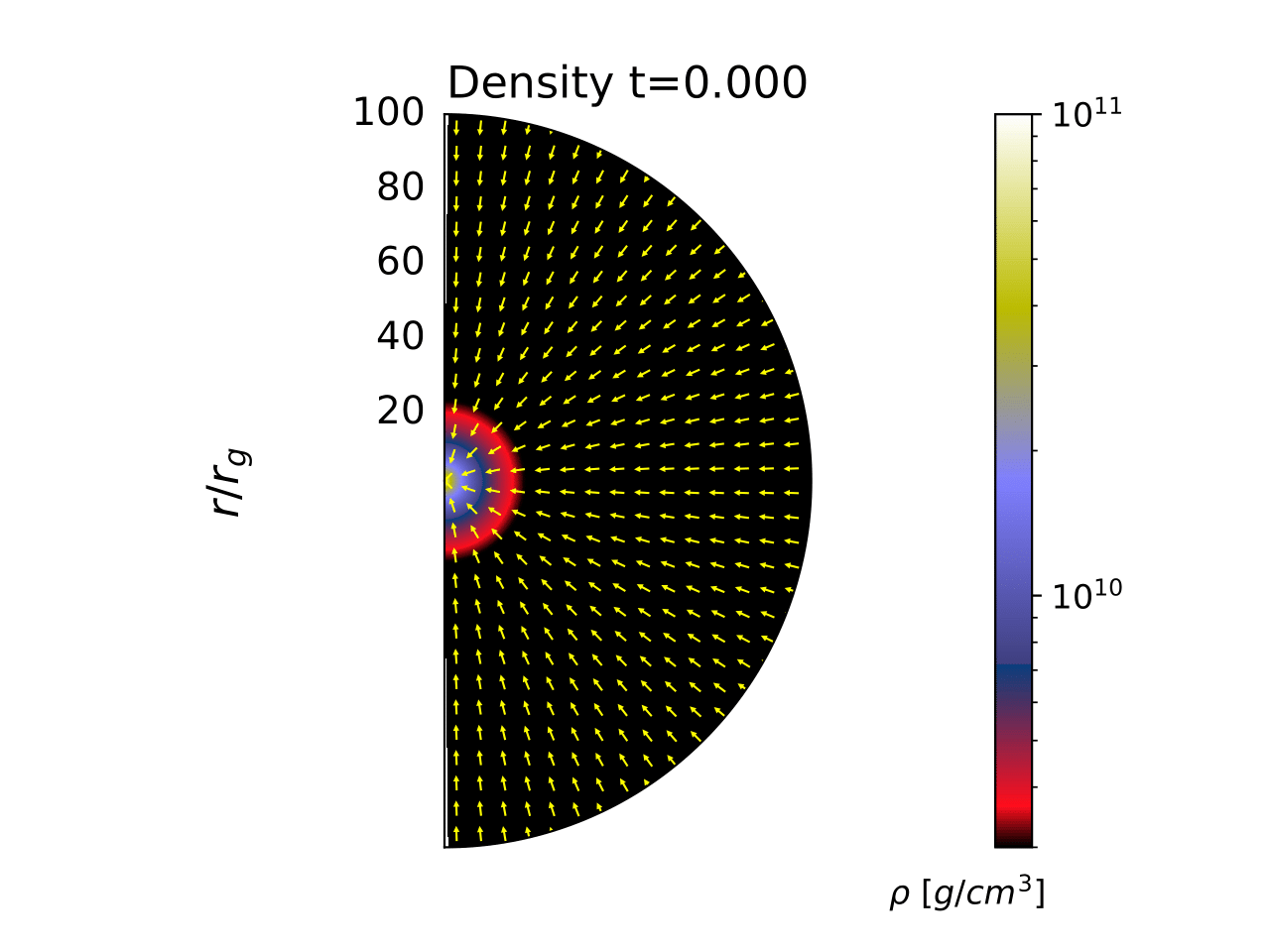}
  \includegraphics[width=0.45\linewidth,height=0.35\linewidth]{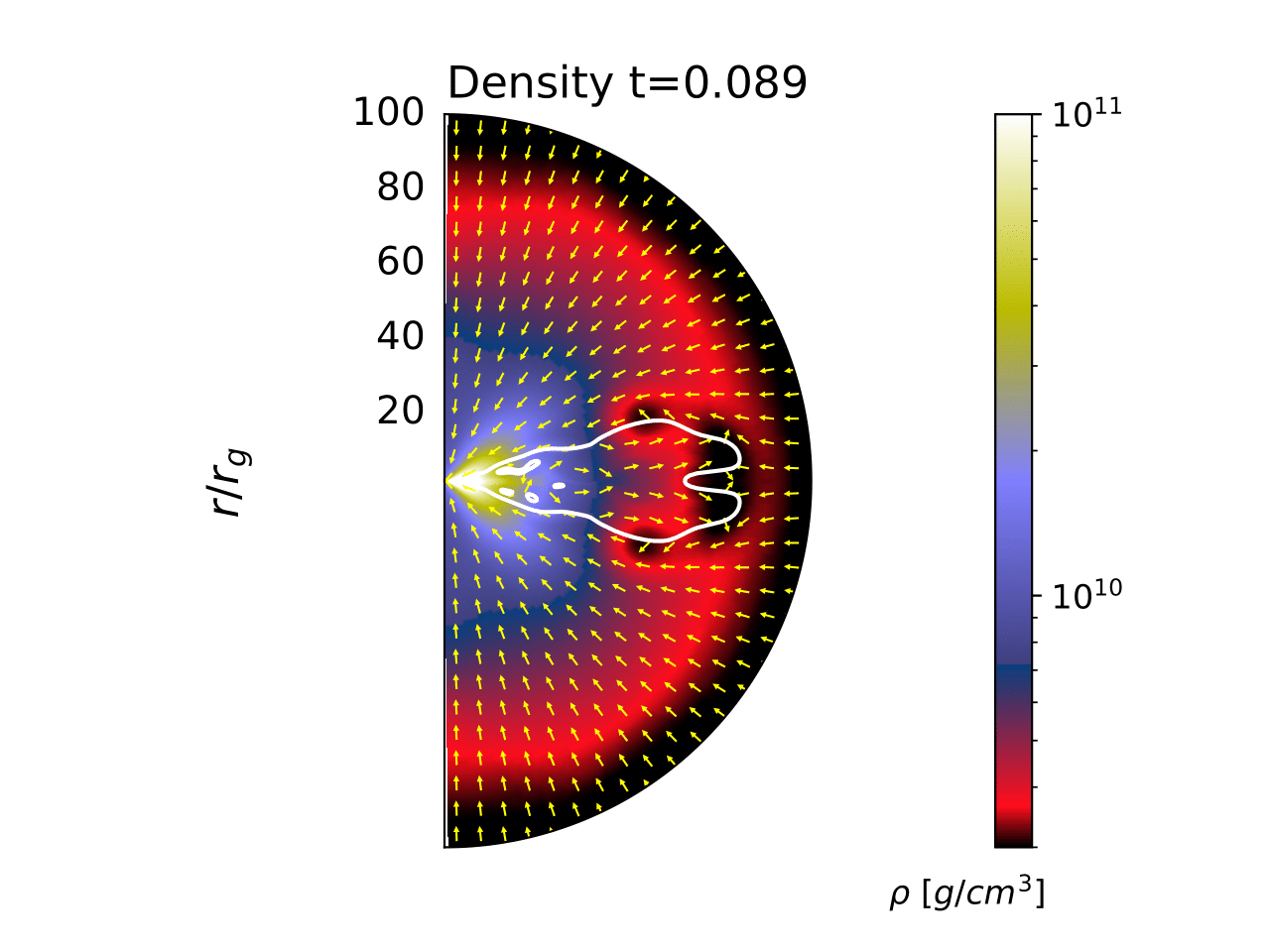}
\end{center}
\caption{\label{Fig:2}
{\bf Left:} Density distribution at the onset of core collapse. Arrows represent velocity field (normalized length). Density distribution and velocity field, after the onset of collapse. Model is parameterized with rotation parameter S=1.4. Thick white line represents sonic surface, Mach=1.
{\bf Right:} Density distribution at the onset of core collapse. Arrows represent velocity field.
}
\end{figure}

We illustrate the initial condition in Figure \ref{Fig:2}, left panel.
Density of the gas is normalized to physical units (given in cgs on the plot), assuming that the collapsing star has the initial mass of 25 Solar masses. This mass is enclosed within our computational domain with outer radius of a $R_{out}=1000 r_{g}$. The plot shows only the innermost region, of 100 $r_{g}$. Most mass of the core is located very near to the center, as it represents the evolved state of stellar evolution with a compact (iron) core formed.

In the initial conditions, we also introduce a small angular momentum imposed on the spherically distributed gas. The specific angular momentum is
normalized by the parameter $S$, with respect to that at the innermost stable circular orbit (ISCO). In addition, the rotation velocity scales with the polar angle, to be maximal at the equator, $\theta=\pi/2$.

   \begin{equation}
l=S l_{\rm isco}r^2\sin^2{\theta}, 
   \end{equation}
   with
   \begin{equation}
l_{\rm isco}=u_{\phi, \rm isco}=\frac{r_{\rm isco}^{1/2}-2a/r_{\rm isco}+a^2/r_{\rm isco}^{3/2}}{\sqrt{1-3/r_{\rm isco}+2a/r_{\rm isco}^{3/2}}}.
   \end{equation}
   
Notice that the radius $r_{\rm ISCO}$ in Kerr geometry depends on the black hole spin. In this proceeding, we show results obtained for the value of initial black hole spin $a_{0}=0.5$. We use several values of rotation parameter, as denoted on the plots in next sections.

After the onset of collapse, the rotation of gas induces formation of a mini-disk, i.e. toroidal structure, located at the equatorial plane. The density distribution becomes no longer spherical. Also, the radial velocity is decreased, as the gas is subject to a centrifugal barrier. Flow is falling into the black hole with supersonic speed from the poles, while at the equator the speed is subsonic.

Map on the Figure \ref{Fig:2}, right panel, shows the flow distribution at time t=0.089 s, for the model normalized with rotation parameter S=1.4. This means that the specific angular momentum is above critical value (S=1) which allows for the formation of rotationally supported torus. Sonic surface, Mach=1, is plotted with a solid line, and marks the location of a transonic shock at the equatorial region.

\section{Impact of Self-Gravity on the collapse}\label{selfg}

In our new simulations, both the mass and angular momentum accreted onto the event horizon ---and used to update the Kerr metric coefficients--- are now modified by the perturbation acting on the metric in the region above the horizon due to the self-gravity force that the gas feels at a given distance from the horizon. These perturbative terms are calculated from the stress--energy tensor. Therefore, in addition to the two equations governing the growth of black hole mass and spin via the mass and angular momentum transfer through the horizon, as given below, \citep{KrolJaniuk2021},
we now add perturbative terms to mass and angular momentum, computed at every radius above the event horizon.

\begin{equation}
 \dot M_{BH} = \int d\theta d\phi\, \sqrt{-g}\, {T^{r}}_t,
\end{equation}
\begin{equation}
\dot{J} = \int d\theta d\phi\, \sqrt{-g}\, {T^{r}}_\phi,
\end{equation} 
\begin{equation}
\delta M_{BH} (t,r) = 2\pi \int_{r_{hor}}^{r} T^{r}_{t}\sqrt{-g} d\theta, 
\end{equation}

\begin{equation}
    \delta J(t,r) = 2\pi \int_{r_{hor}}^{r}T^{r}_{\phi}\sqrt{-g} d\theta, 
\end{equation}
\begin{equation}
    \delta a = {{J + \delta J(r)} \over {M_{BH}+\delta M_{BH}(r)}} - a^{i}, 
\end{equation}
\begin{equation}
a^{i}=a^{i-1}+\Delta a.
\end{equation}

The terms computed in addition to mass and angular momentum changes
\citep{Janiuk2018}
as these self-gravity perturbations, are integrated at each grid point in the radial direction and at each time.
They affect the change of Kerr metric coefficients, which are sensitive to the mass and spin updates.
The dimensionless black hole spin, a, evolves as a result of black hole mass and angular momentum changes due to accretion of mass under the horizon, and is additionally changed due to self-gravity of the collapsing core. The numerical
method has been described in detail in \cite{Janiuk2023}.
Below, we compare the results of self-gravitating collapsar models to the
runs without self-gravity, in order to emphasize the difference and to investigate the role of self-gravity in the collapsar physics.

\begin{figure}[p]
\begin{center}
  \includegraphics[width=0.3\linewidth,height=0.2\linewidth]{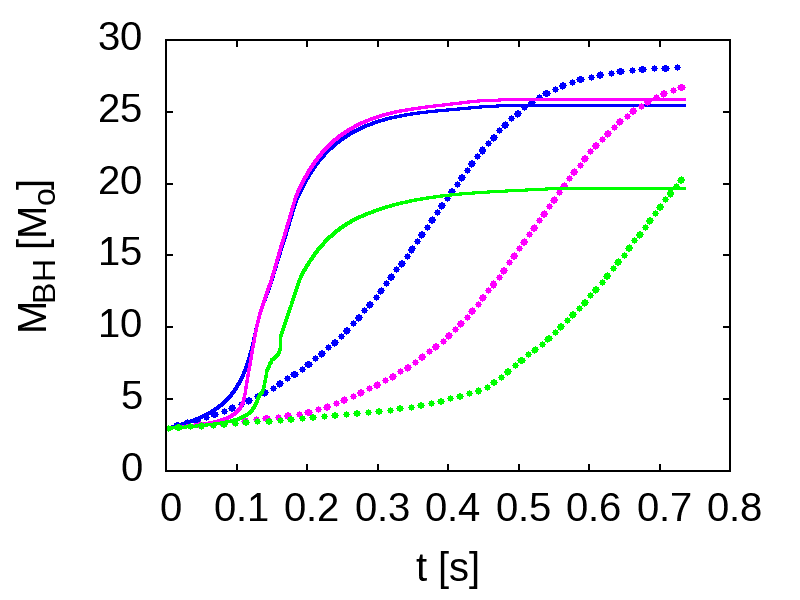}
  \includegraphics[width=0.3\linewidth,height=0.2\linewidth]{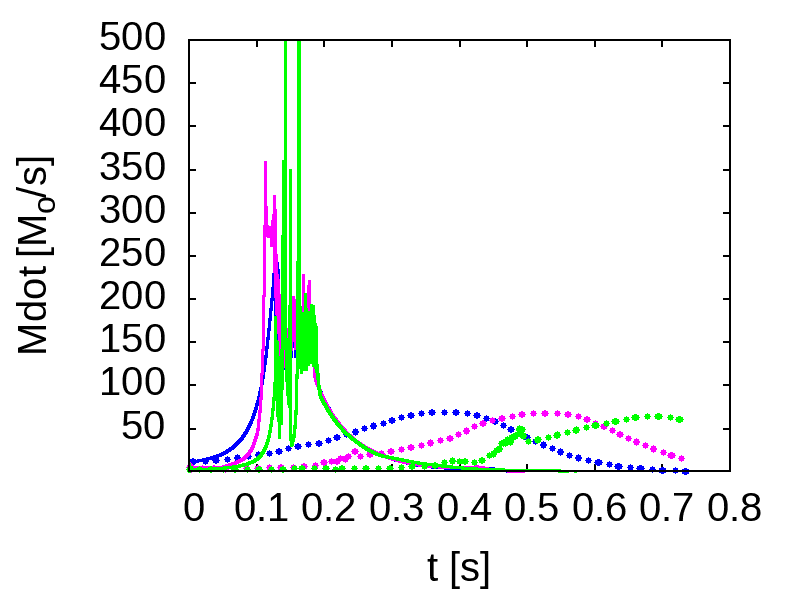}
  \includegraphics[width=0.3\linewidth,height=0.2\linewidth]{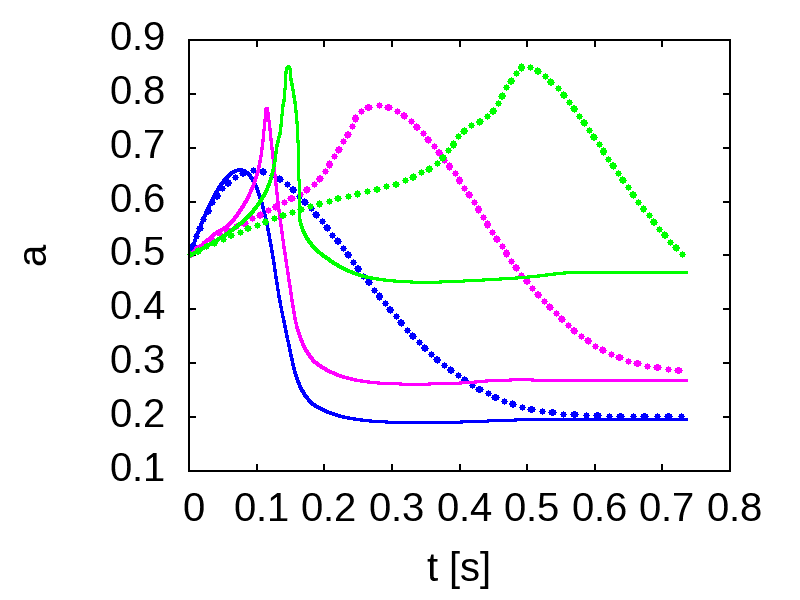}
\end{center}
\caption{\label{Fig:3}
  {\bf Left:} Black hole mass changing as function of time during the collapse. We start from 3 Solar mass black hole. Models including and excluding self-gravity are plotted by the thick and thin curves, respectively. Three color refer to different amount of angular momentum in the collapsing star: S=1.0 (blue), S=1.4 (red) and S=2.0 (green).
  {\bf Middle:} Evolution of accretion rate through the horizon, for self-gravitating and non-self-gravitating collapsars, shown with thick and thin lines, respectively. The different colors refer to various amounts of angular momentum in the collapsar, same as in the left plot.
  {\bf Right:} Evolution of the black hole dimensionless spin parameter, during the collaspe. We start from oderately spinning black hole with a=0.5. Models including and excluding self-gravity are plotted by the thick and thin curves, respectively. Three color refer to different amount of angular momentum in the collapsing star, same as in the left and middle plots.
}
\end{figure}

As shown in Figure \ref{Fig:3}, the results are strongly sensitive to the adopted self-gravity effects, and also weakly sensitive to the rotation of the collapsing envelope. The latter is normalized with respect to the critical angular momentum, for which the flow is circularized at the innermost stable orbit, ISCO \citep{KrolJaniuk2021}.
In addition, the rotation velocity scales with the polar angle, so that at the equator, the rotation of the star is maximal. We notice that the larger the initial rotation magnitude, the longer it takes for the black hole mass to evolve. The non-SG simulations end with very different final black hole mass, depending on the rotation parameter.

In contrast, the self-gravity of the envelope can speed up the evolution of the collapsing stellar core significantly. Also, accretion rate and its fluctuations are of much higher amplitude when self-gravity effects taken into account. Without self-gravity, there are longer time intervals where there is considerably less fluctuation of the accretion rate; in this case, there exist only some small oscillations in the accretion rate during some time intervals (around 0.2 s for S=1.4, and 0.4-0.5 s for S=2).

\subsection{Instabilities on the collapsing core}\label{RTI}

\begin{figure}[p]
\begin{center}
  \includegraphics[width=0.3\linewidth,height=0.25\linewidth]{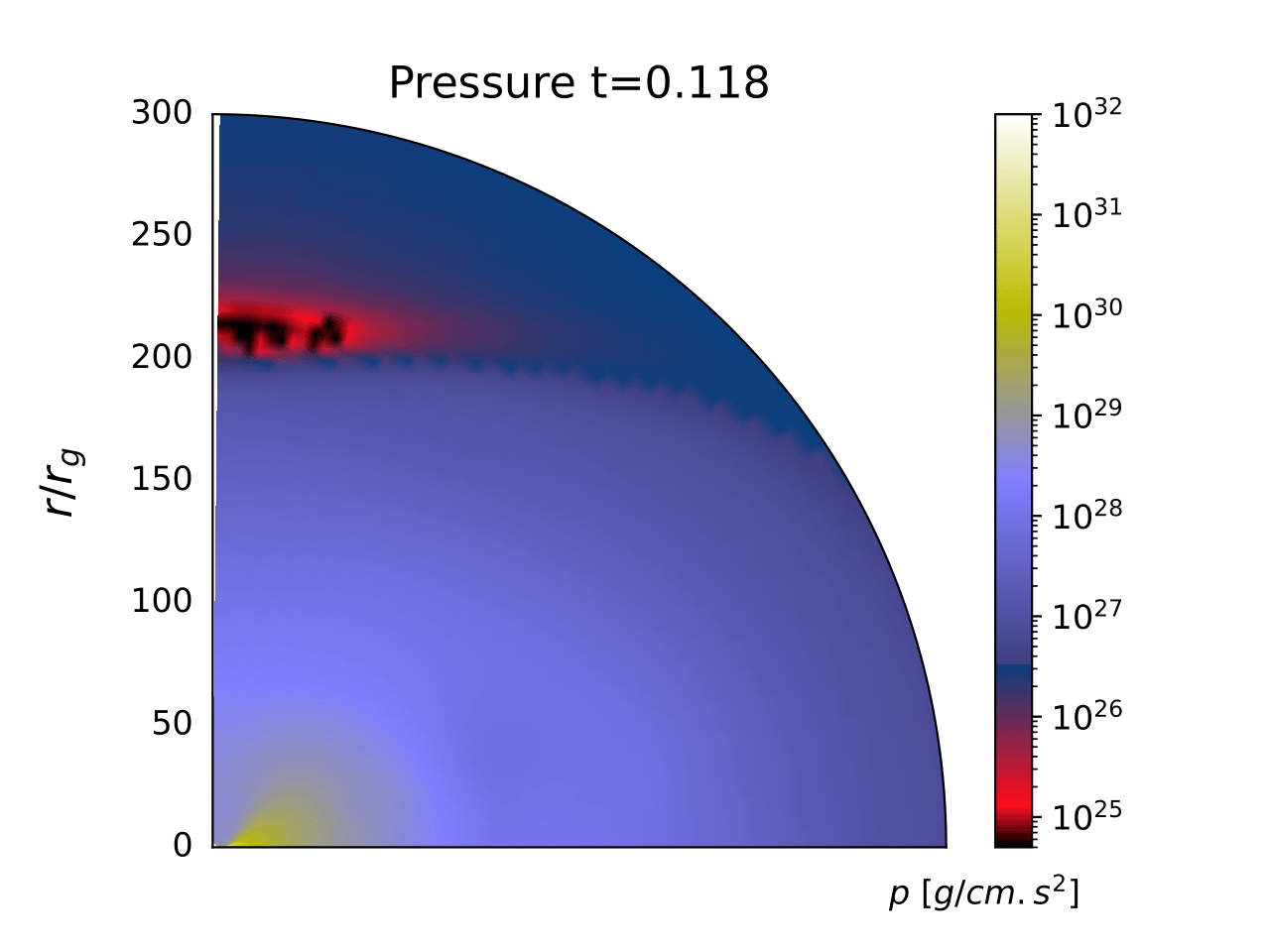}
  \includegraphics[width=0.3\linewidth,height=0.25\linewidth]{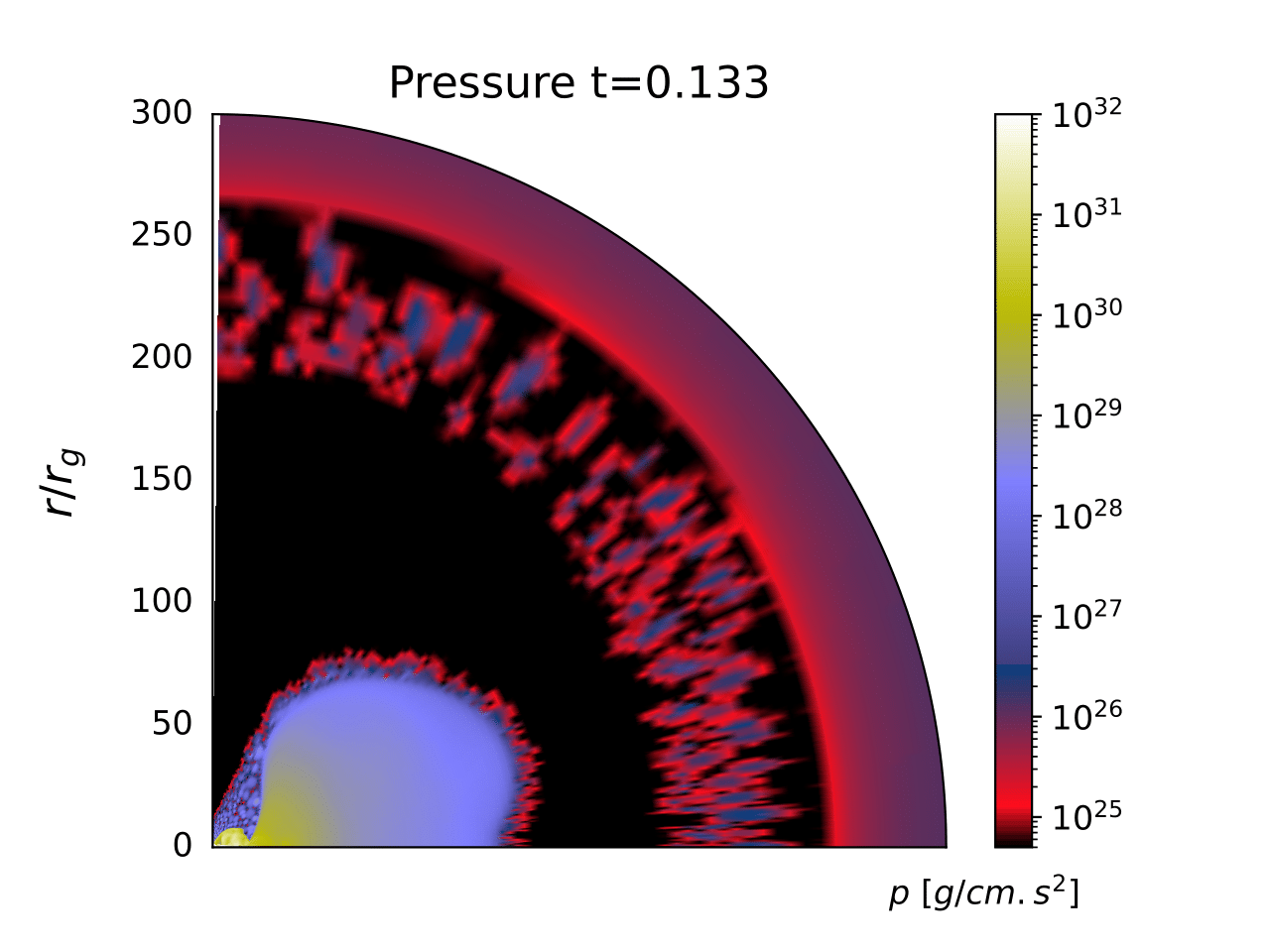}
    \includegraphics[width=0.3\linewidth,height=0.25\linewidth]{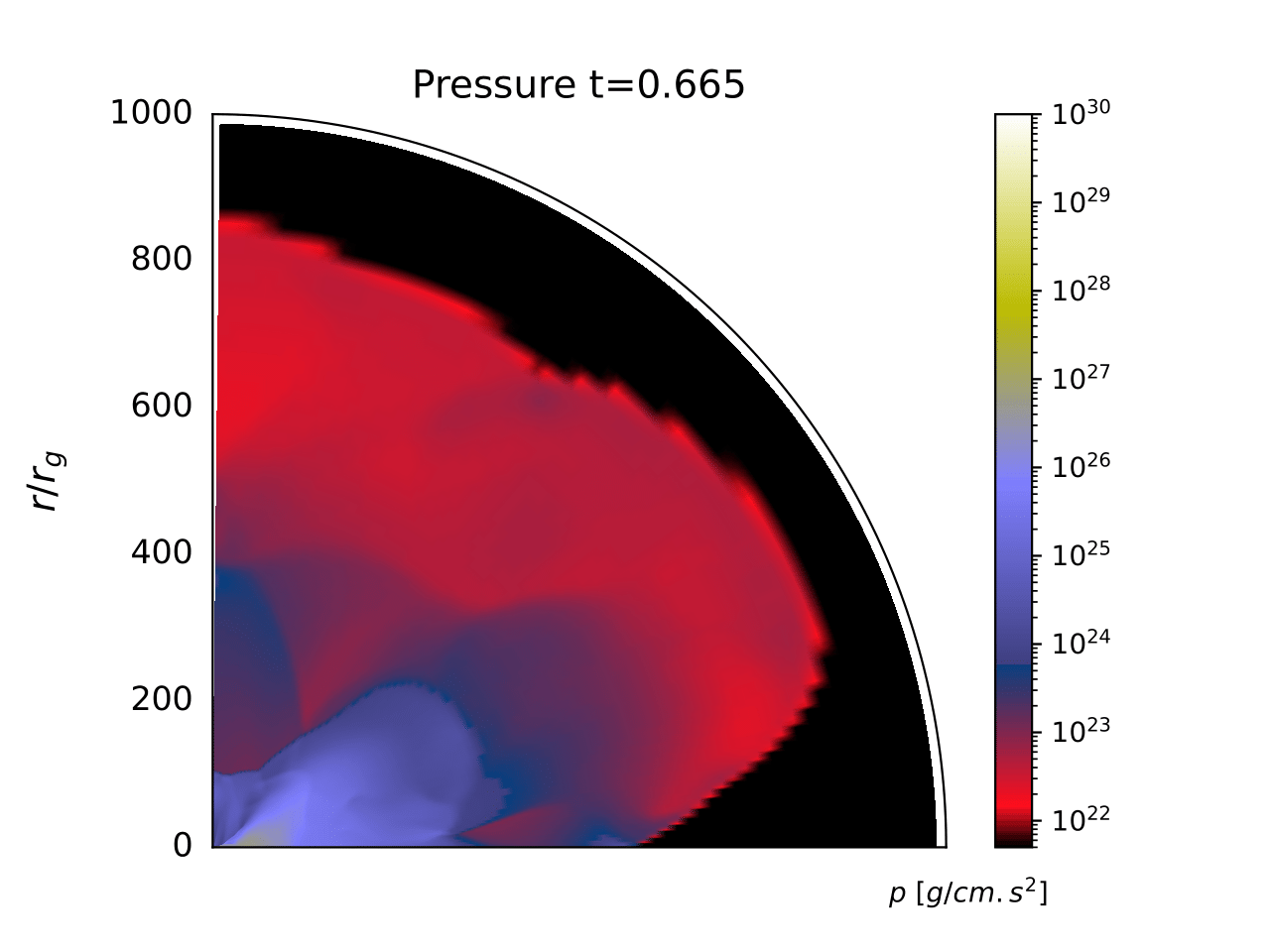}
\end{center}
\caption{\label{Fig:4}
  {\bf Left:} Pressure profile for the model with rotation parameter S=2 and initial black hole spin $A_{0}=0.5$, taken at time t=0.118 s., at which largest accretion rate fluctuations appear.
  {\bf Middle:} Pressure profile at time t=0.133, for the same model as in the left plot. Strong inhomogeneity regions are visible. 
{\bf Right: } Pressure profile at later time of the simulation, for the same model as in the left and middle plots. Inhomogeneities are same and at this time accretion rate fluctuations are smoothed as well. The map is zoomed out to larger radius.
}
\end{figure}

\begin{figure}[p]
\begin{center}
  \includegraphics[width=0.3\linewidth,height=0.25\linewidth]{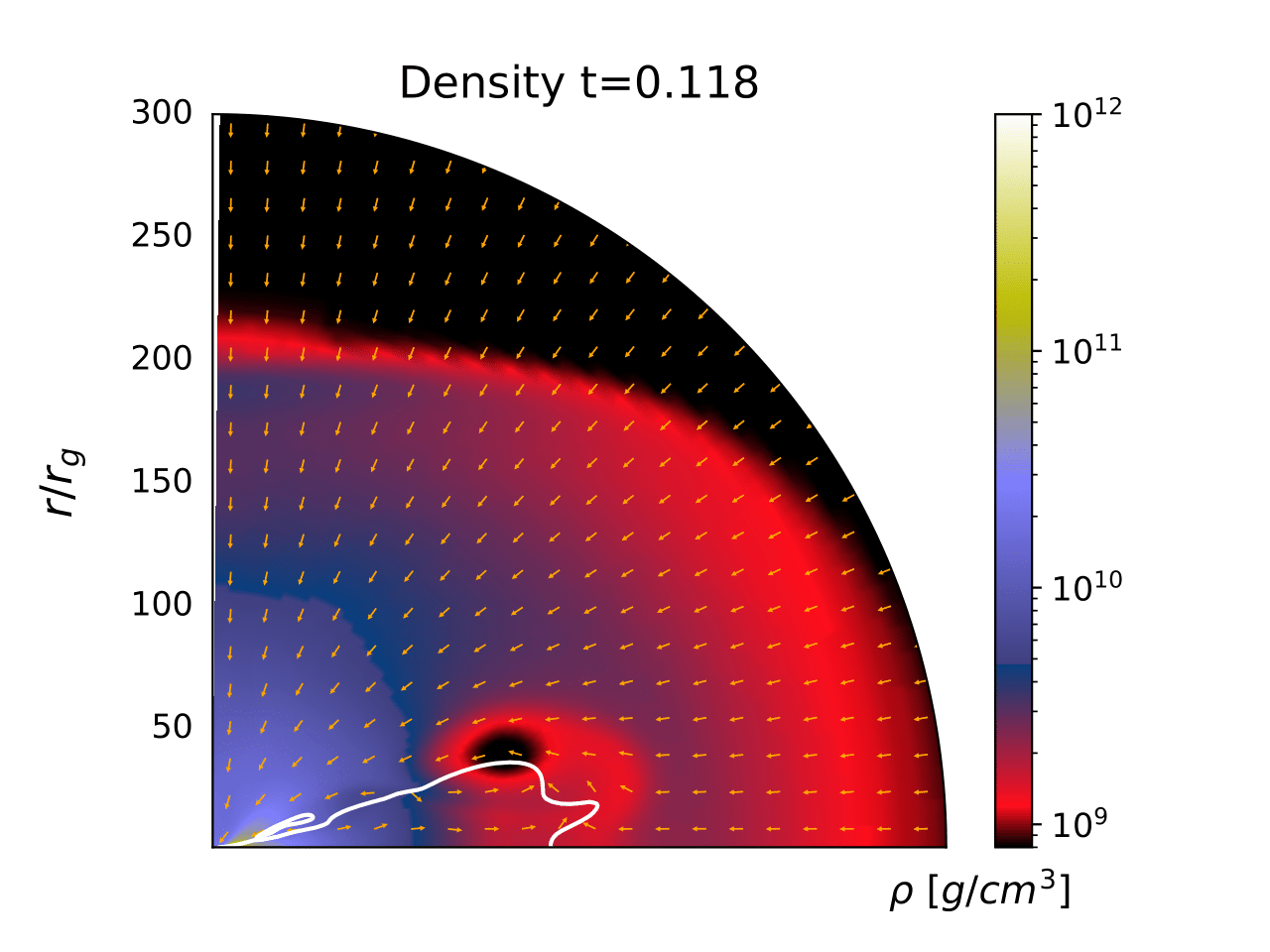}
  \includegraphics[width=0.3\linewidth,height=0.25\linewidth]{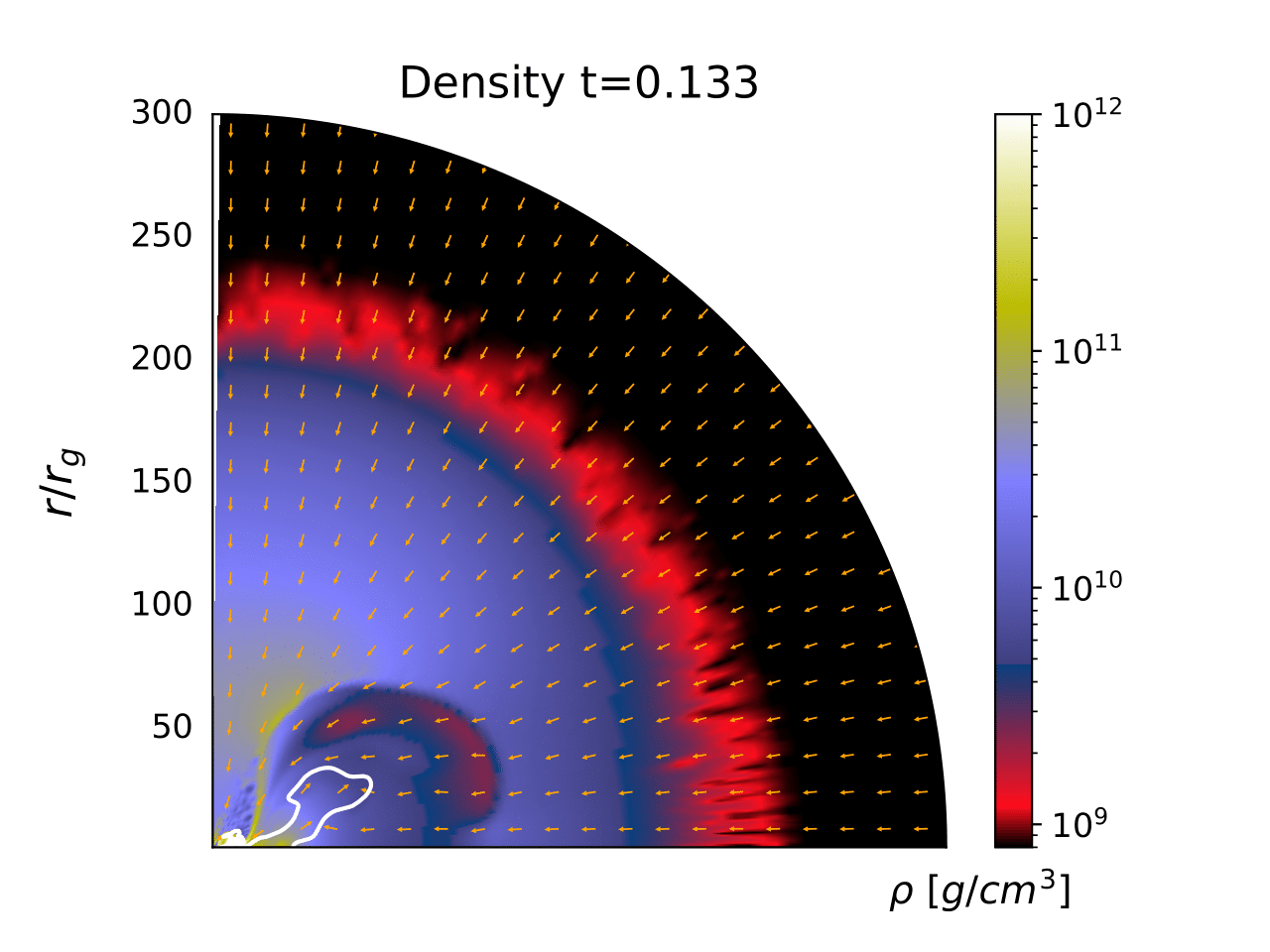}
    \includegraphics[width=0.3\linewidth,height=0.25\linewidth]{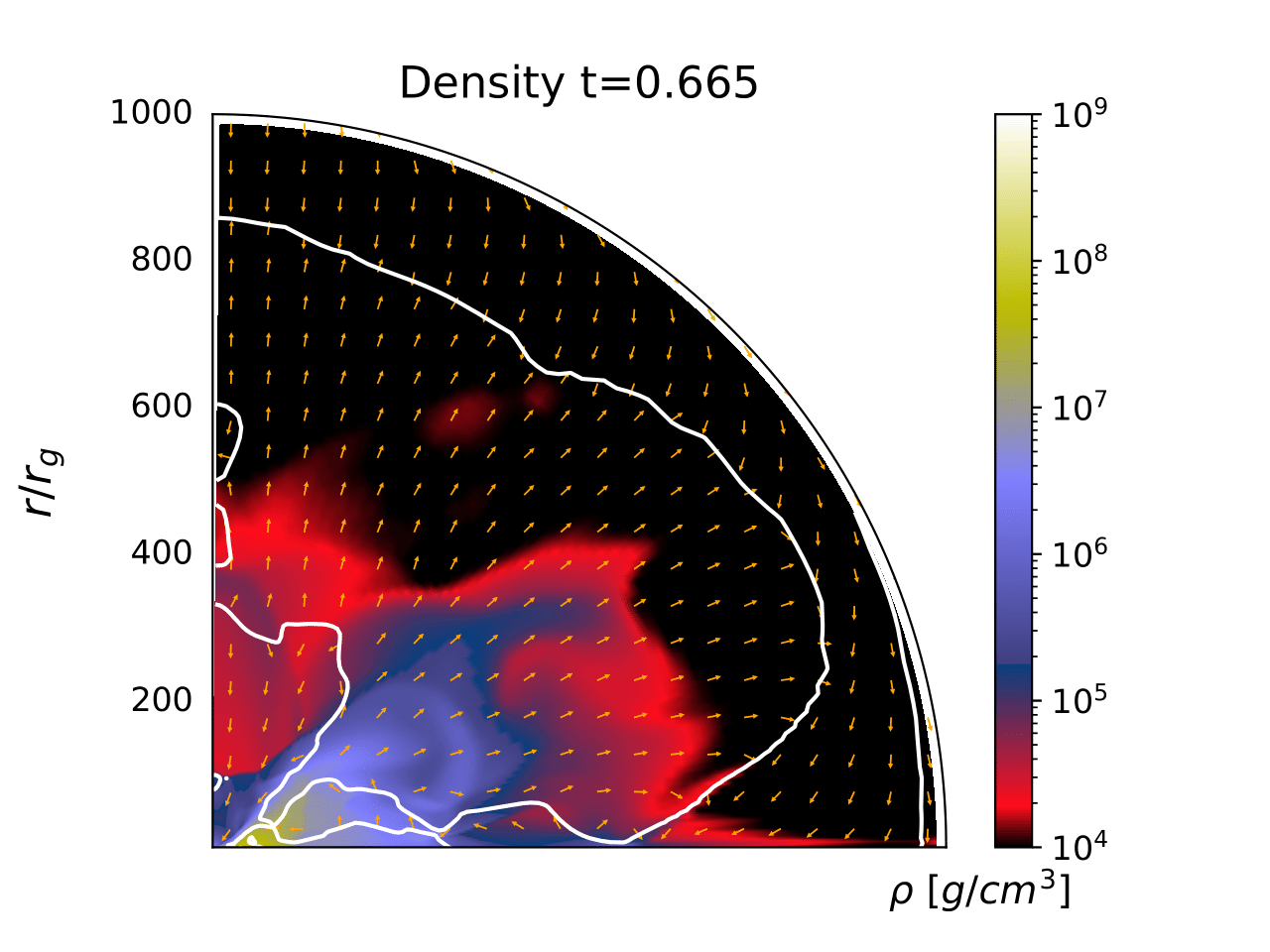}
\end{center}
\caption{\label{Fig:5}
  {\bf Left:} Density profile at t=0.118, for the same model as above.
  {\bf Middle:} Density profile at t=0.133, for the same model as above.
{\bf Right:} Density profile at late time, t=0.665, for the same model as above.
}
\end{figure}

As an effect of self-gravity we observe density inhomogeneities and formation of the accretion shocks in all our models, regardless of the initial black hole spin, or rotation parameter of the collapsar. First, there appears an equatorial outflow of matter, which reaches radii of up to about 80 $r_{g}$ and is then stalled in the transonic shock. The small inhomogeneities in the pressure and density at the chosen time intervals, are visible in more detail in the plots below, in Figure \ref{Fig:4} and in Figure \ref{Fig:5}, respectively.

We quantify the inhomogeneities in the collapsar by computing the radial derivatives of density an pressure at specific times, and locations. We identify the mechanism for their creation as the SGI instability (Self-Gravity Interfacial instability) and we compare its strength with another well-known hydrodynamical instability, the Rayleigh-Taylor (RT) instability. Their growth rates are given as below \citep{kifonidis2003non, hunter1997kelvin}.

\begin{equation}
\sigma_{RT} = \sqrt{-\frac{p}{\rho}\frac{\partial ln \rho}{\partial r}\frac{\partial ln p}{\partial r}},
\end{equation}
\begin{equation}
\sigma_{SGI} = \sqrt{\frac{2\pi G (\rho_{2}-\rho_{1})^{2}}{(\rho_{2}+\rho_{1})}}.
\end{equation}

The RT and SGI instabilities result in very similar configurations at density snapshots. However, they have their own characteristics, which allows us to differentiate between them. As self-gravity has no `preferred' direction, it is destabilizing across all density interfaces, while an interface is RT-unstable only if the heavy fluid is on top of the light fluid. It has also been confirmed that RT instability is characterized by dense spikes penetrating the tenuous fluid, whereas the SGI develops with tenuous spikes streaming into the denser fluid.

We find that SGI instability seems to dominate over RT instability and produces the inhomogeneities. In particular, we checked that the growth rates of RT, are having imaginary values, as computed at radii between 20 and 25 $r_{g}$, around the mixing boundary.

\begin{figure}[p]
\begin{center}
  \includegraphics[width=0.45\linewidth,height=0.25\linewidth]{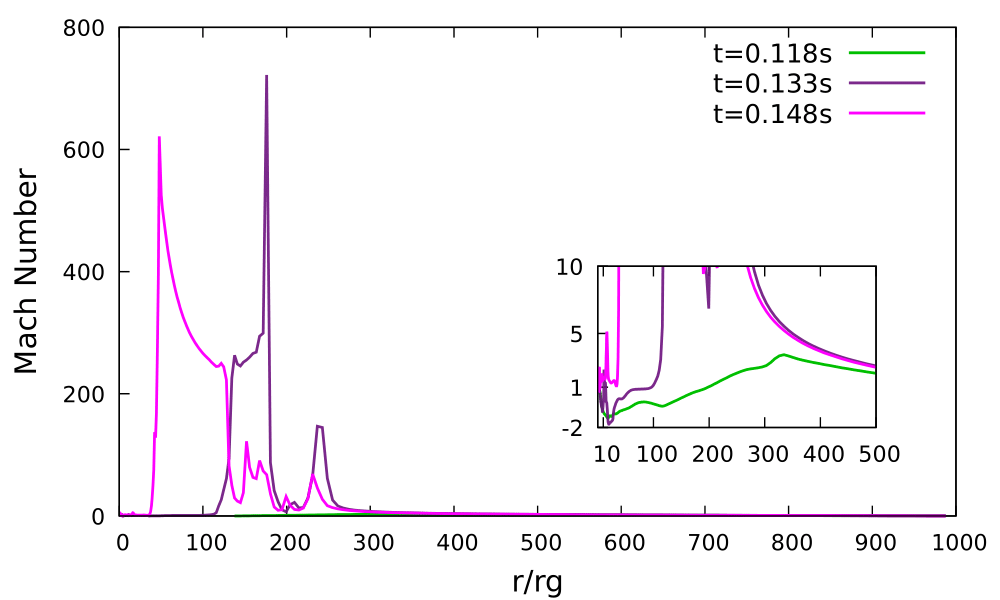}
    \includegraphics[width=0.45\linewidth,height=0.25\linewidth]{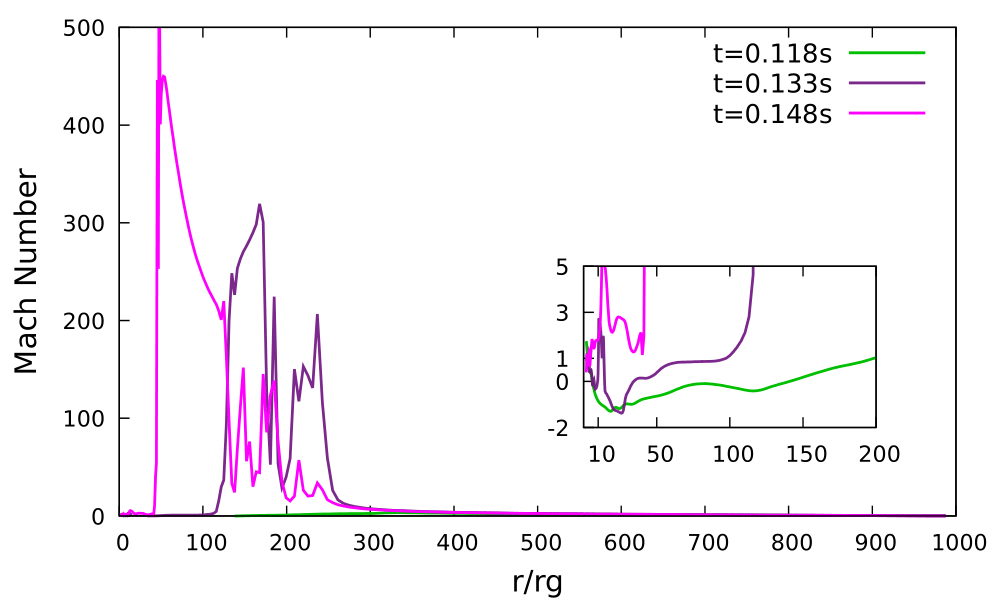}
\end{center}
\caption{\label{Fig:6}
  {\bf Left:} Mach number profile at three different times,
  t=0.118, t=0.133, and t=0.148 s, for the model with $S=2.0$.
  {\bf Right:} Mach number profile at three different times, for the same model with magnetic field.
}
\end{figure}

Finally, we investigated the formation of transonic shocks in the collapsars. In Figure \ref{Fig:6} we present radial profiles of Mach number at some specific time snapshots, for models with $S=2$ and $a_{0}=0.5$. The left panel shows the profiles in the self-gravitating case, while the plot in the right panel shows those of self-gravitating magnetized case (we introduced a weak vertical magnetic field in the initial condition).
For the sake of more visibility, we provide zommed-in inset panels representing the inner regions. 
We observe the sonic front expansion, and also some transient shock formation during the collapse.
At early times, the small transonic shocks appear around 100 $r_{g}$ and they present a moderate density contrast (pre-shock to post-shock density ratio $R=\rho_{1}/\rho_{2}\sim 10$). Such shocks  also appear at later times. Their formation is enhanced by the self-gravity effects. 
We find that magnetic field does not make any significant difference on the shock expansion timescales, but it affects the strength of the shock, consistently with previous studies \citep{1999MNRAS.308.1069K}.

\section{Conclusions}\label{conclus}

In this work, we show numerical models of the collapsing stellar core where we  account for the dynamical evolution of central black hole mass and its spin. The related coefficients of the Kerr space-time metric are evolved accordingly, at every time step. In addition, we calculate the self-gravity of the stellar envelope and we add the relevant perturbative terms to the dynamical evolution of the black hole spin parameter.

The last modification of the model turned out to have an impact on the global evolution of the collapsing star, and produces dramatic fluctuations in the
accretion rate at the initial phase of collapse.
More importantly, it also plays crucial role in development of the SGI interfacial instability in its specific regions. We identified inhomogeneities in density and pressure distributions which arise due to self-gravity, and we concluded that the SGI instability dominates over the RT, as its growth rate is positive in the regions of mixing boundaries.


\ack

The present work was supported by the grant
DEC-2019/35/B/ST9/04000 from Polish National Science Center. We made use of computational resources of the PL-Grid infrastructure, under grant pglgrb6, and Warsaw University ICM.
D. Ł. K. was supported by the Polish National Science Center Dec-2019/35/O/ST9/04054 and N. Sh. D. was supported by Iran National Science Foundation (INSF) under project number No.4013178 and also acknowledges Ferdowsi University of Mashhad (FUM), Iran, and the FUM Sci-HPC center. Prof. Shahram Abbassi also deserves gratitude for his accompaniment to N.Sh.D. in this project. A.J. acknowledges the Czech-Polish mobility program (MŠMT 8J20PL037 and PPN/BCZ/2019/1/00069).

\bibliography{ragajaniuk}

\end{document}